\documentclass[epj]{webofc}
\usepackage[varg]{txfonts}   
\usepackage{graphicx}

\usepackage{hyperref}
\usepackage{float}
\usepackage{epsfig}
\usepackage{graphicx}
\usepackage{mathptmx}      
\usepackage{latexsym}
\usepackage{longtable}
\usepackage{dcolumn}
\usepackage{natbib}
\usepackage{comment}
\usepackage{placeins}
\newcommand{\tr}{{\rm tr\,}}
\newcommand{\Tr}{{\rm tr\,}}

\newcommand{\eins}{\leavevmode\hbox{\small1\kern-3.8pt\normalsize1}}
\newcommand{\be}{\begin{eqnarray}}
\newcommand{\ee}{\end{eqnarray}}

\wocname{EPJ Web of Conferences}
\woctitle{CONF12}
\begin{document}
\selectlanguage{english} 
\title{Complex Langevin simulations of a finite density matrix model for QCD}
%
%

\author{Jacques Bloch \inst{1} \and 
 Jonas Glesaaen \inst{2} \and Owe Philipsen\inst{3} \and 
Jacobus Verbaarschot \inst{4} \and Savvas Zafeiropoulos \inst{5} \fnsep\thanks{Presenter; \email{savvas@jlab.org}.} }
\institute{Institute for Theoretical Physics, University of Regensburg, 93040 Regensburg, Germany\\ \and Department of Physics, College of Science, Swansea University,
Swansea, SA2 8PP, United Kingdom\\ \and
Institut f\"ur Theoretische Physik, Goethe-Universität Frankfurt \\
Max-von-Laue-Str. 1, 60438 Frankfurt am Main, Germany\\ \and
Department of Physics and Astronomy\\
Department of Physics and Astronomy, Stony Brook University, Stony Brook, NY, 11794  \\ \and 
Jefferson Laboratory, 12000 Jefferson Avenue, Newport News, VA 23606, USA\\
Department of Physics, College of William and Mary, Williamsburg, VA 23187-8795, USA }

\abstract{We study a random matrix model for QCD at finite density via complex Langevin dynamics. This model has a phase transition to a phase with nonzero baryon density. We study the convergence of the algorithm as a function of the quark mass and the chemical potential and focus on two main observables: the baryon density and the chiral condensate. For simulations close to the chiral limit, the algorithm has wrong convergence properties when the quark mass is in the spectral domain of the Dirac operator. A possible solution of this problem is discussed.}

\maketitle
\section{Introduction}\label{intro}
The study of the phase diagram of QCD in the temperature ($T$)-chemical potential
($\mu$) plane is one of the biggest challenges in Quantum Chromodynamics.
There is a plethora of ab-initio nonperturbative studies employing lattice
simulations at finite temperature and zero baryon density which provided many
celebrated results. However, due to the infamous sign problem, studies at
$\mu\neq 0$ are seriously impeded. Finite baryon chemical potential for
$SU(3)$ QCD with fundamental quarks leads to a loss of $\gamma_5$-hermiticity and consequently a complex fermion determinant. As a result one can not use powerful
Markov Chain Monte Carlo methods since the probability weight is not positive
definite. There have been many workarounds in the literature trying to avoid the
direct confrontation with the sign problem. The most prominent lattice approaches
are reweighting, Taylor expansions of thermodynamic observables around
$\mu=0$ and imaginary chemical potential simulations.
Unfortunately these suffer from their own problems, namely the overlap problem
and a limited radius of convergence which limit us to study only the regime
of $\mu/T<1$, cf.~\cite{Owe} for a pedagogical review. It is thus clear that one needs to develop new methods in order
to exit from this deadlock. One possible solution could be employing the
rather old method of stochastic quantization that was first introduced more
than thirty years ago independently by Parisi and Klauder~\cite{Parisi, Klauder}.
This method was abandoned soon after its introduction
because, although stochastic quantization reproduces the results of path
integral quantization in the case when the action is real~\cite{PHD},
this is not necessarily the case when the action is complex. The main problem, which
is quite worrisome, is that, in some cases, we have convergence to the wrong solution~\cite{AartsET}.

In recent years, the complex Langevin (CL) method was revived in a series
of papers by Aarts and collaborators \cite{Aarts:2009uq,AartsET}.
Since then the method has reproduced correct results for
many non-trivial theories and toy models and
it seems to be working fine for QCD in the deconfined phase as well
as with heavy quarks~\cite{sexty1, sexty2}.
The question to which the whole community would like to know the answer is
if CL will work correctly close to the chiral limit in the confined phase. This was, for example, recently investigated in two-dimensional strong-coupling QCD \cite{{Bloch:2015coa}}, where it was shown that CL leads to wrong results for small masses. 

In this contribution, we will address this question by studying a random matrix theory (RMT) model
of QCD at nonzero chemical potential which has both a complex fermion determinant
and a finite density phase transition. This model 
was first introduced by Stephanov in~\cite{Misha} and has been extensively
scrutinized. This is not the first CL study employing a matrix model for QCD --
there has been extensive work~\cite{KS1, KS2, J1} on a finite density
model for QCD introduced by Osborn~\cite{James} and improved in
Ref.~\cite{Bloch-etal}
which however does not have
a phase transition and could only serve as a model for QCD for small values of the chemical potential.
RMT has played a crucial role in providing analytical understanding
for non-perturbative aspects of QCD related to finite density as well as in improving
the understanding of the effect of discretization errors and topology on the
Dirac spectrum~\cite{DSV, ADSV, KVZprl, KVZprd, KVZ2c, CSZ, pdrmt, like}.
In this work we are studying the convergence properties of the CL algorithm
for a model of continuum QCD at nonzero chemical potential.
The model has a known analytic solution, which gives us a handle on the
discretization errors otherwise appearing due to lattice artefacts.

\section{The Stephanov model}
We study a random matrix model for QCD at finite density proposed by Stephanov~\cite{Misha} whose partition function reads
 \begin{equation}
 \mathcal{Z}^{N_f}_n=e^{N\mu^2}\int dX dX^{\dagger}{\det}^{N_f}(D+m)e^{-N\,\Tr XX^\dagger}.
 \label{Zst}
 \end{equation}
  The Dirac operator $D$ has the form
 \begin{equation}
  D=\left(
 \begin{array}{cc}
 m        & iX+\mu\\
 iX^\dagger+\mu & m
 \end{array}
 \right),
 \label{Dsteph}
\end{equation}
where we have added the chemical potential term $\mu\gamma_0$ to the RMT Dirac
operator \cite{ShVe93}. The $N\times (N+\nu)$ matrix elements of $X$ are
independently distributed complex Gaussian random variables, $N$ is the size of
the block matrix $X$ and $\nu$ is the index of $D$. Therefore, in accordance
with the Atiyah-Singer index theorem, $\nu$ can be identified as the topological
charge of the gauge fields.
We set $\nu=0$ throughout the article as the topological charge has no, or
little, effect on the main subject of this study.
As is the case for QCD, the RMT has a serious sign problem when the quark mass is inside the spectral domain of the Dirac operator.

The two main observables that we consider are the mass dependent chiral condensate   $\Sigma(m)=\langle\bar{\eta}\eta\rangle$
and the baryon number density $n_B=\langle \eta^{\dagger}\eta\rangle$
defined as 
\begin{equation}
\langle\bar{\eta}\eta\rangle=\frac 1{2N}\frac{\partial \log{ \mathcal{Z}^{N_f}_N}}{\partial m},
\label{PBP}
\end{equation}
and
\begin{equation}
\langle\eta^{\dagger}\eta\rangle=\frac 1{2N}\frac{\partial \log{ \mathcal{Z}^{N_f}_N}}{\partial \mu}.
\label{nb}
\end{equation}
As already mentioned, one can incorporate the chemical potential $\mu$ in several ways, some of which have particular advantages~\cite{James}. For example,
in the Osborn model~\cite{James}, another candidate matrix model,   one can obtain the joint probability distribution function with standard RMT techniques.
Consequently, one can construct all  $k$-point eigenvalue correlation functions. This model is a schematic model for QCD at finite chemical potential $\mu$ but does not have the same rich phase structure as the Stephanov model, which possesses a phase transition to a phase with nonzero baryon density. One needs to bear in mind that all these models result in the same chiral Lagrangian as long as they share the same global symmetries. Note that the chiral Lagrangian for QCD with three colors does not depend
on the chemical potential.
In~\cite{HJV} the unquenched partition function of the Stephanov model is reduced analytically to a form that can easily be evaluated numerically at finite $N$, or
can be solved completely by a saddle point approximation  in the thermodynamic limit where $N\to\infty$.  The $N_f=1$ partition function, in units where the
chiral condensate $\Sigma=1$, can be cast via bosonization in the following $\sigma$-model form
 \be
\mathcal{Z}^{N_f=1}_N(m,\mu) = e^{N\mu^2}\int d\sigma d\sigma^* e^{-N\sigma^2} (\sigma \sigma^* +
  m(\sigma+\sigma^*) + m^2 -\mu^2)^N \ ,
  \label{zoneflavor}
 \ee
 where $\sigma$ is the bosonized form of $\bar{\psi}_L \psi_R$. If one changes to polar coordinates, the angular integral can be performed analytically and yields a modified Bessel function. The partition function is then given by the following one-fold integral
 \be
\mathcal{Z}^{N_f=1}_N(m,\mu) = \pi e^{-Nm^2+N\mu^2}\int_0^\infty du
 (u-\mu^2)^N I_0(2mN\sqrt u) e^{-Nu} \ .
 \label{zi0}
\ee
The strategy that we adopt is very clear. We will simulate the Stephanov model via CL and will verify its convergence properties by comparing to analytical results for the chiral condensate and the baryon density using the partition function
(\ref{zi0}).

\section{Complex Langevin}\label{CL}
In order to briefly recall the basics of stochastic quantization  we consider the trivial "QFT" given by the partition function
$\mathcal{Z}=\int e^{-S(x)} dx$ .
 The real Langevin algorithm is given by  
\be
x(t+\delta t)=x(t)-\partial_x S(x(t))\delta t +\delta \xi 
\ee where the stochastic variable $\delta \xi$ has zero mean and variance
$2\delta t$. Generalization to complex actions is rather trivial, as one simply
promotes each dof to their complex counterpart. This is necessary since the
drift term ($\partial_x S(x(t))\delta t$) is complex and pushes the variables
into the complex plane. Naturally $x$ gets replaced by $z=x+iy$ 
which will evolve according to
\be
z(t+\delta t)=z(t)-\partial_z S(z(t))\delta t +\delta \xi.
\ee

Adapting this formalism for RMT is quite straightforward. In a cartesian representation, the complex random matrix is decomposed as $X=a+ib$ and $X^{\dagger}=a^{\top}-ib^{\top}.$ 
The integration measure is $dX dX^{\dagger}=da db$ and the action reads
\be
S=N\tr(X^{\dagger}X)-N_f \tr(\log(m^2-\mu^2 +X^{\dagger}X-i\mu (X+X^{\dagger}))).
\ee
Then one defines the matrix G by
\be
G=(m^2-\mu^2+X^{\dagger}X -i\mu (X+X^{\dagger}))^{-1}.
\ee
\label{action}
so that the Langevin update steps for the matrices $a$ and $b$ read
\be
a^{(n+1)}_{mn}=a^{(n)}_{mn}-2N\delta t a_{mn}+N_f \delta t [(XG)_{mn}+(X^*G^{\top})_{mn}-i\mu (G_{mn}+G_{mn}^{\top})]+\delta \xi ,
\label{cartesianAupd}
\ee
\vspace{-6mm}
\be
b^{(n+1)}_{mn}=b^{(n)}_{mn}-2N\delta t b_{mn}+N_f \delta t [(XG)_{mn}-(X^*G^{\top})_{mn}-i\mu (G_{mn}^{\top}-G_{mn})]+\delta \xi .
\label{cartesianBupd}
\ee
Note that in the above expressions $X^{*}=a-ib$ since the complexification has not taken place yet. 

\begin{figure}[t!]
\centerline
{
\begin{minipage}{0.46\textwidth}
\centerline{%
\hspace{0.1cm}\includegraphics[width=1.07\linewidth]{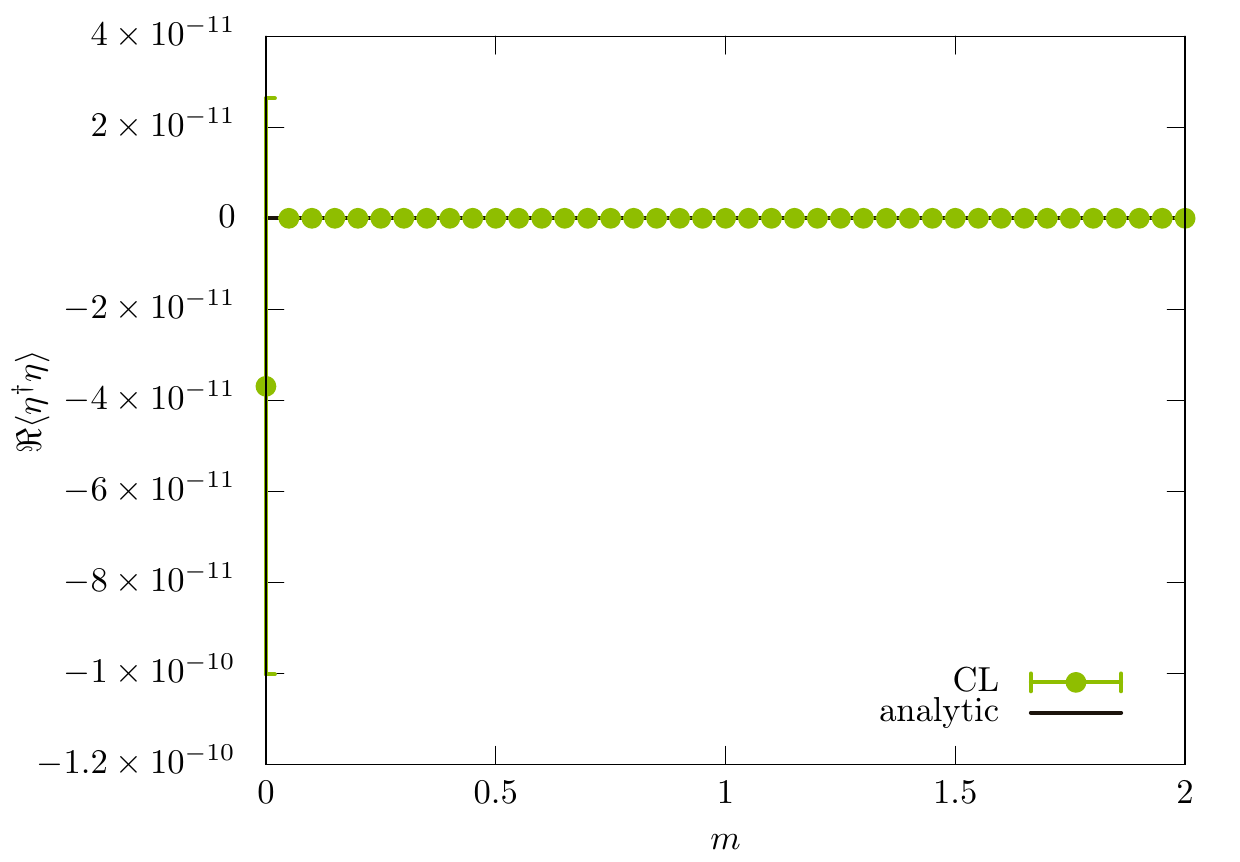}}
\end{minipage}
\hspace{5mm}
\begin{minipage}{0.46\textwidth}
\centerline{%
\hspace{0.5cm}\includegraphics[width=1.07\linewidth]{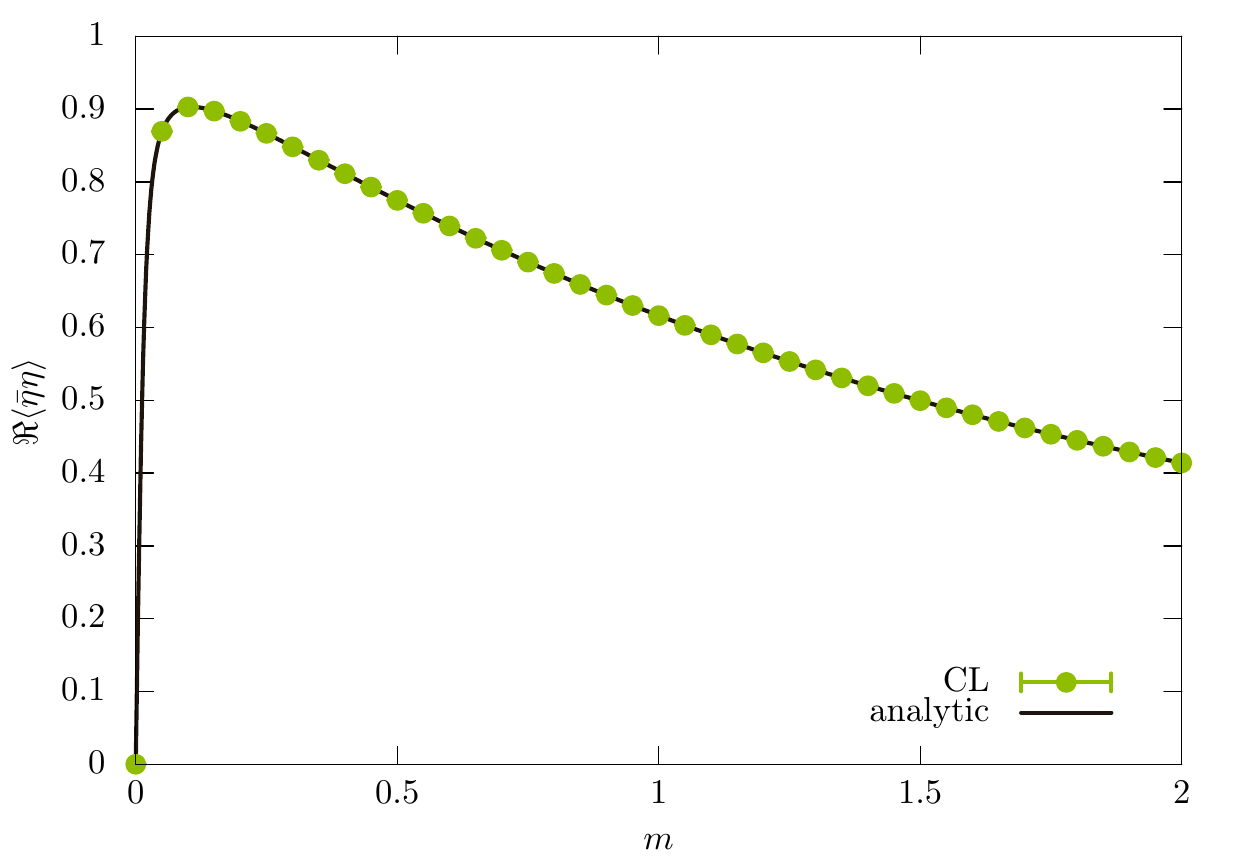}}
\end{minipage}
}
\caption{\label{numsanity0}$\langle \eta^{\dagger}\eta\rangle$ (left) and $\langle \bar{\eta}\eta\rangle$  (right), both as a function of $m$ for $\mu=0$.}
\end{figure}

\begin{figure}[htb]
\centerline
{
\begin{minipage}{0.46\textwidth}
\centerline{%
\hspace{0.1cm}\includegraphics[width=1.07\linewidth]{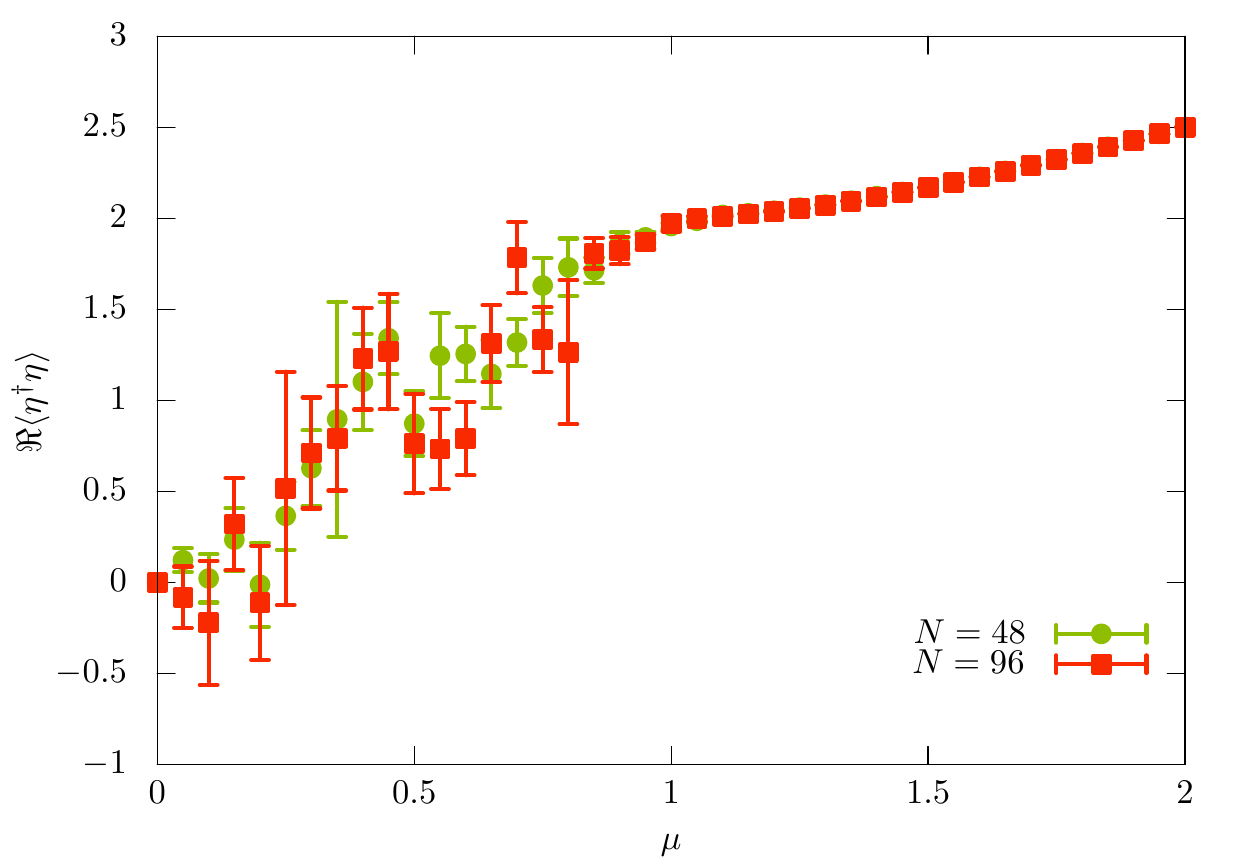}}
\end{minipage}
\hspace{5mm}
\begin{minipage}{0.46\textwidth}
\centerline{%
\hspace{0.5cm}\includegraphics[width=1.07\linewidth]{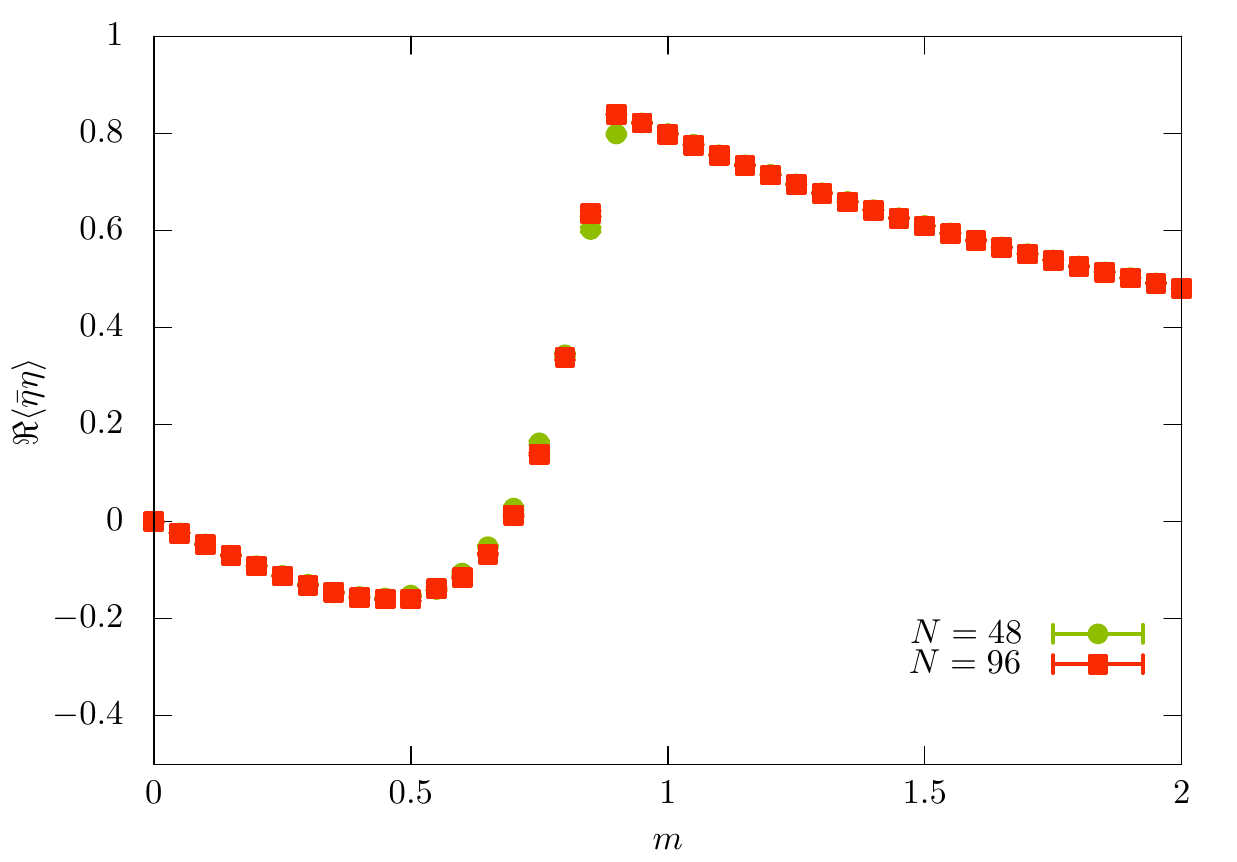}}
\end{minipage}
}
\caption{Matrix size sensitivity: \label{numsanity}$\langle \eta^{\dagger}\eta\rangle$ versus $\mu$ for $m=0$ (left) and $\langle \bar{\eta}\eta\rangle$ versus $m$ for $\mu=1$ (right).}
\end{figure}

\begin{figure}[htb]
\centerline
{
\begin{minipage}{0.46\textwidth}
\centerline{%
\hspace{0.1cm}\includegraphics[width=1.07\linewidth]{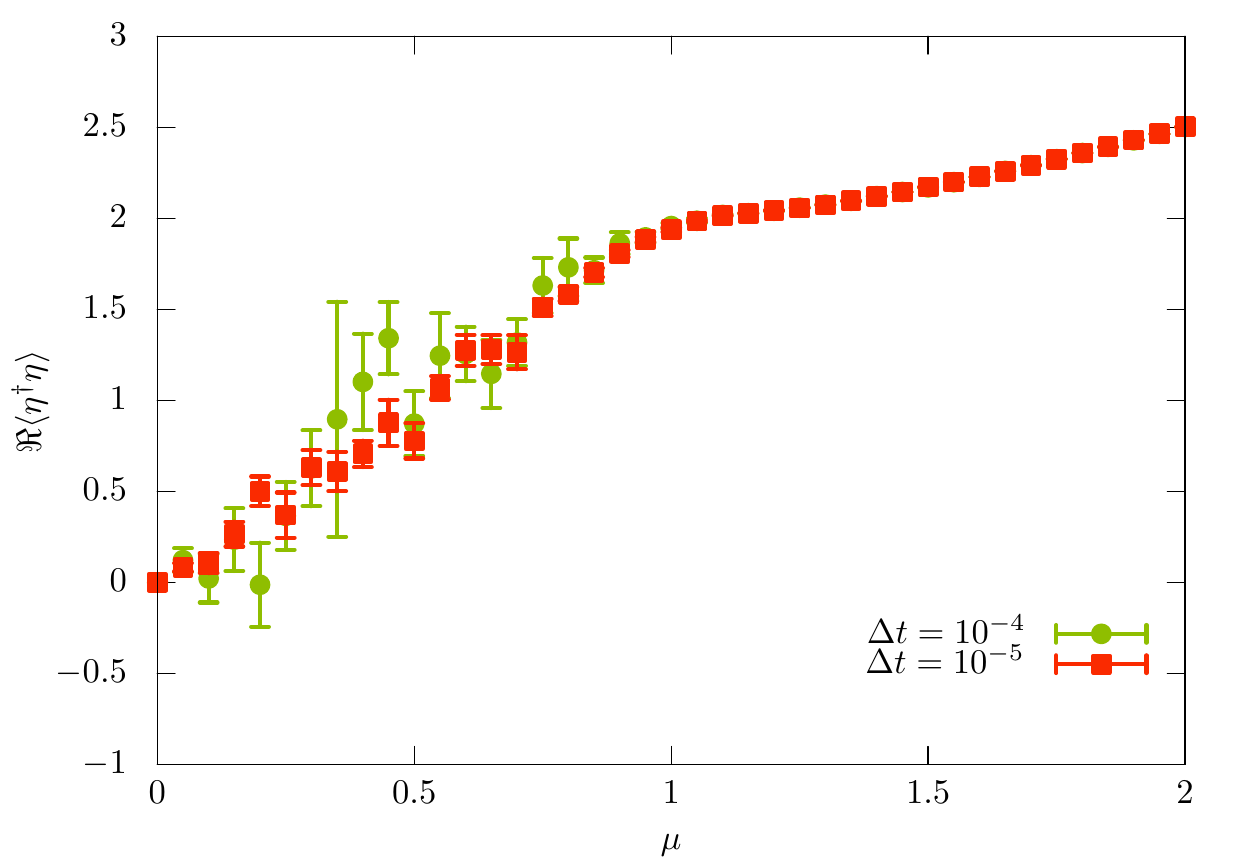}}
\end{minipage}
\hspace{5mm}
\begin{minipage}{0.46\textwidth}
\centerline{%
\hspace{0.5cm}\includegraphics[width=1.07\linewidth]{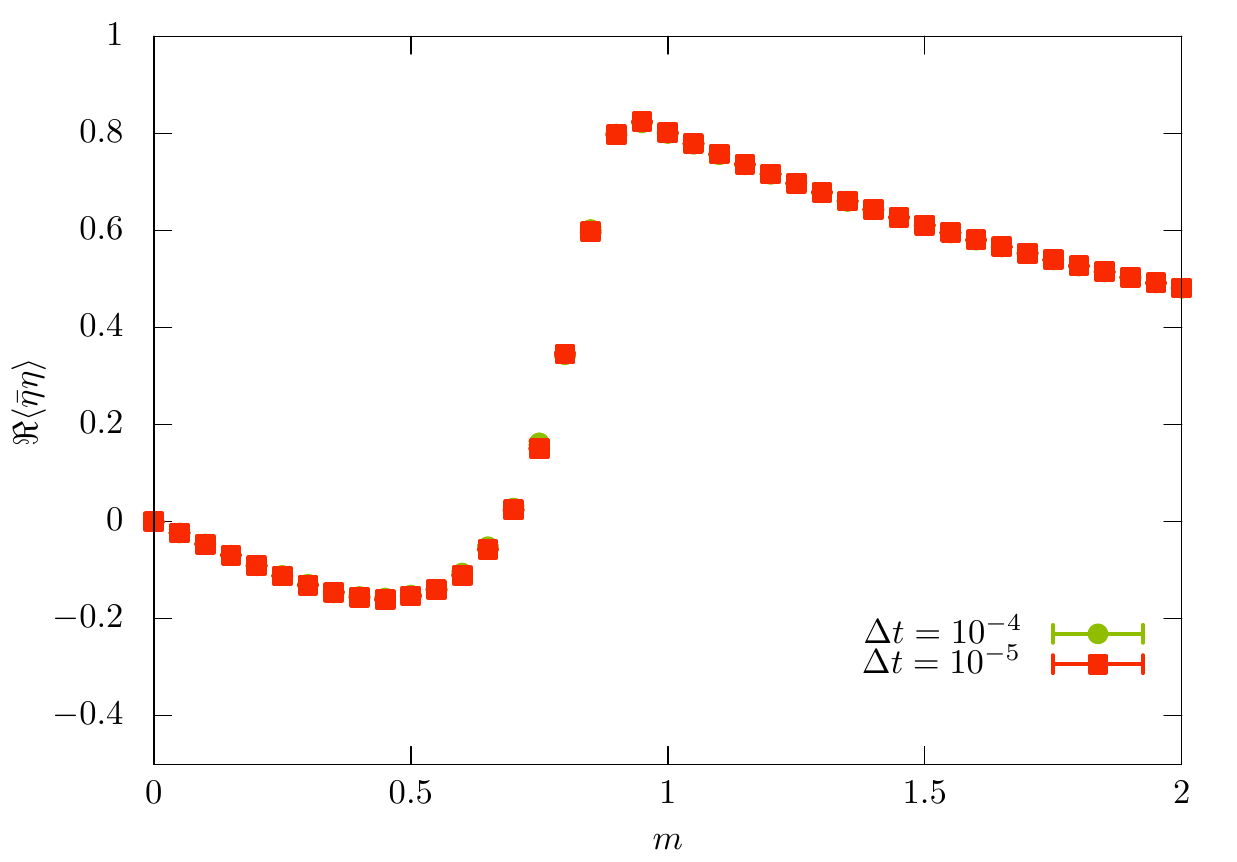}}
\end{minipage}
}
\caption{Step size sensitivity: \label{numsanity2} $\langle \eta^{\dagger}\eta\rangle$ versus $\mu$ for $m=0$ (left) and $\langle \bar{\eta}\eta\rangle$ versus $m$ for $\mu=1$ (right).}
  \end{figure}

The first checks that one needs to perform is that the numerical simulation gives the correct answer in the case of a real action (when $\mu = 0)$, and one needs to verify numerically that there is no sensitivity on the matrix size as well as the step size of the integration. These tests are shown in Figs. \ref{numsanity0}, \ref{numsanity} and \ref{numsanity2}.

At nonzero chemical potential
we perform two types of scans in the space of parameters. In the first type we fix the chemical potential and scan with respect to the quark mass, while in the second type we fix the quark mass and scan with respect to the chemical potential. Starting with the mass scans we readily see the following behavior. For relatively heavy quark mass and small values of the chemical potential the algorithm reproduces the correct result but it has troubles to do so close to the chiral limit as one can see in the upper panel of Fig.~\ref{mscan}. The discrepancy between the numerical and
analytical results widens for increasing values of the chemical potential.
It is noteworthy that the algorithm converges to the results for the phase quenched theory, which is given by the red curve in Fig.~4.
One comment is in order at this point. For the dynamical theory we have obtained
analytical results for finite matrix size $N$ and we can compare that with the
numerical simulation for the same value of $N$. The phase quenched theory, on
the other hand, is analytically more complicated and we give the mean field
result obtained through a saddle point approximation. At finite $N$ the sharp
kinks of the phase quenched results would have been smoothened out.

\FloatBarrier



We now turn to scans at fixed quark mass where one can check if the algorithm
reproduces the phase transition separating (in the chiral limit) a phase with
zero baryon density from a phase with nonzero baryon density. In Fig.\ \ref{muscan} we show the results for three different masses starting with the chiral case. 
In RMT,
one can actually perform a cost effective simulation directly in the chiral limit and one immediately observes that the algorithm especially struggles in this case. However the agreement becomes better as one increases the quark mass.

  In all cases one observes that the algorithm  converges to the phase quenched theory,
  and in the cases where it agrees with the full dynamical simulations, it is
  clearly for the regimes where the phase quenched and the full theory coincide. The next question that we would like to address is if there is hope to actually rectify this pathological behavior. We will make an attempt in this direction in the next section.

\begin{figure}[ht] 
  \label{ mscan} 
  \begin{minipage}[b]{0.5\linewidth}
    \centering
    \includegraphics[width=.95\linewidth]{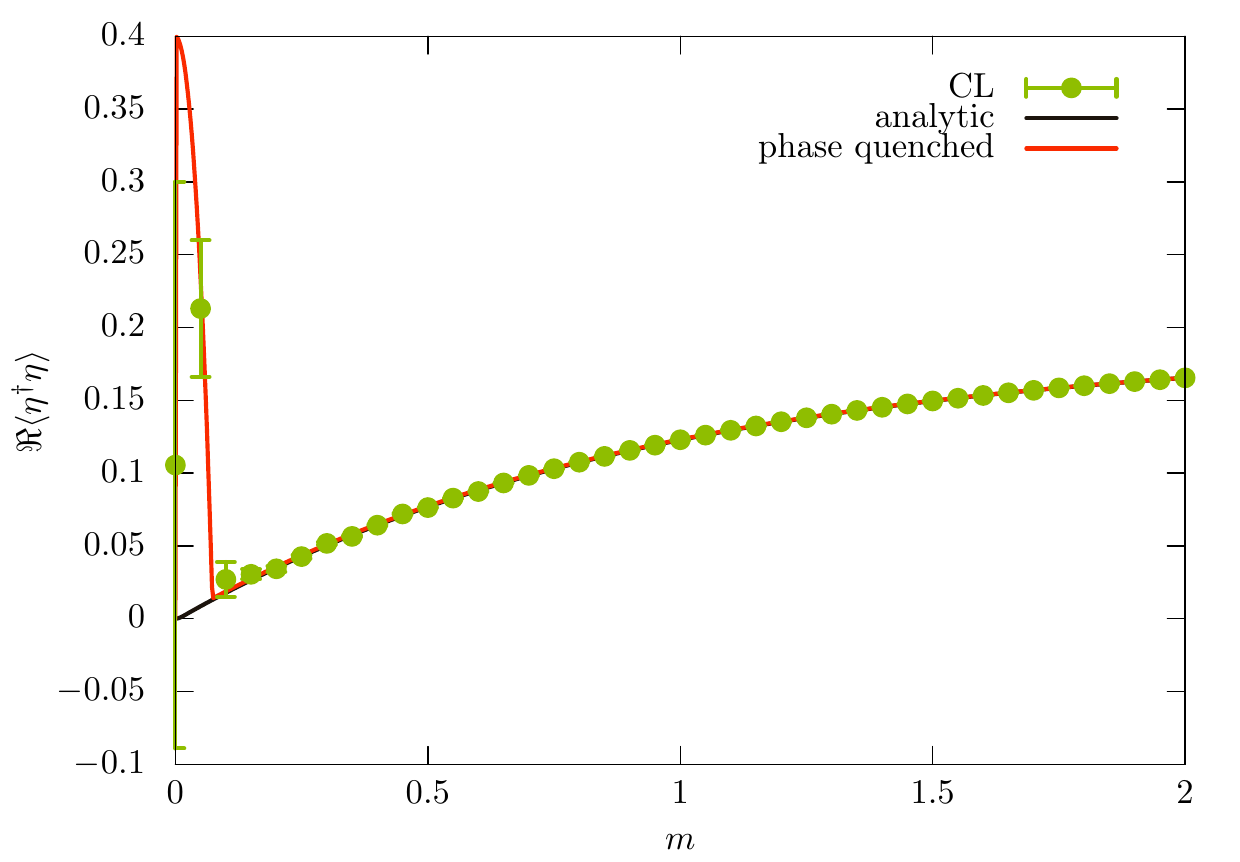} 
    \vspace{4ex}
  \end{minipage}
  \begin{minipage}[b]{0.5\linewidth}
    \centering
    \includegraphics[width=.95\linewidth]{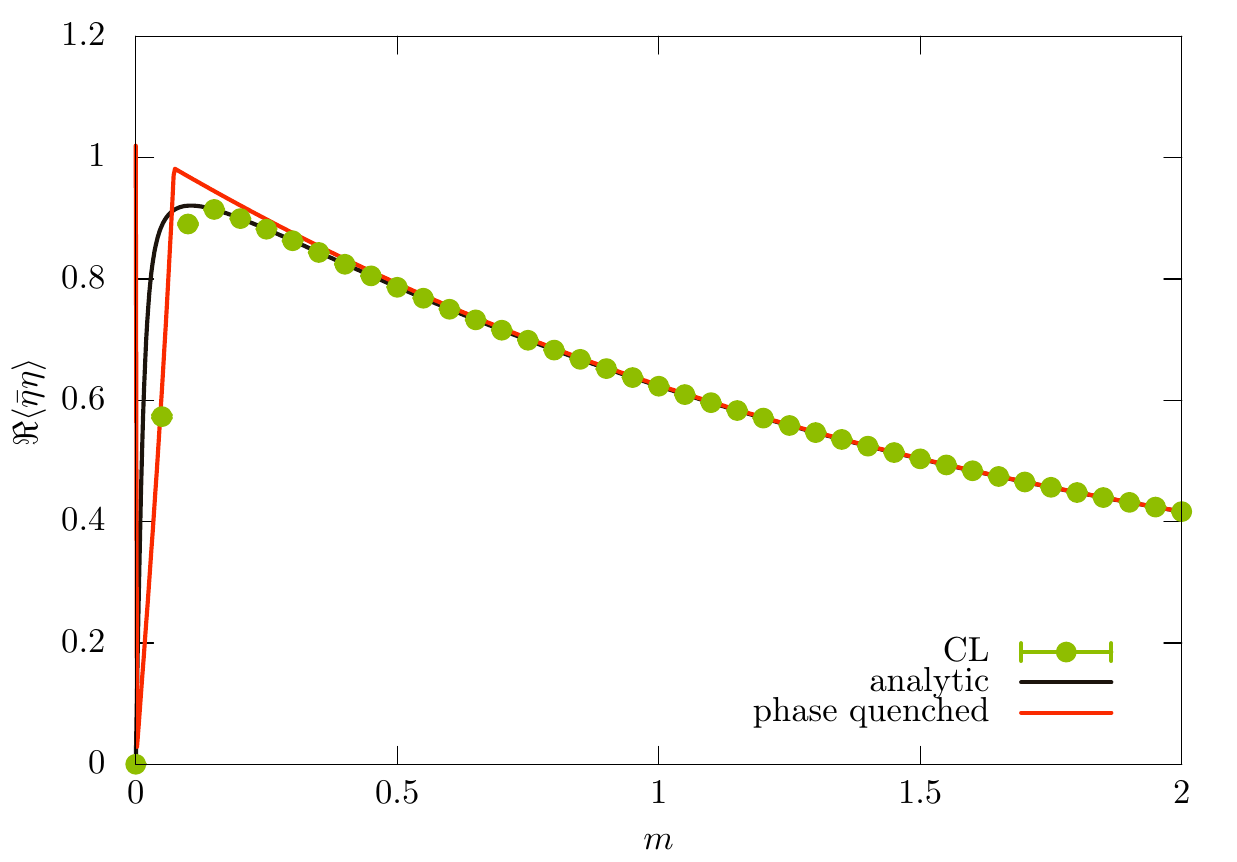} 
    \vspace{4ex}
  \end{minipage} 
  \begin{minipage}[b]{0.5\linewidth}
    \centering
    \includegraphics[width=.95\linewidth]{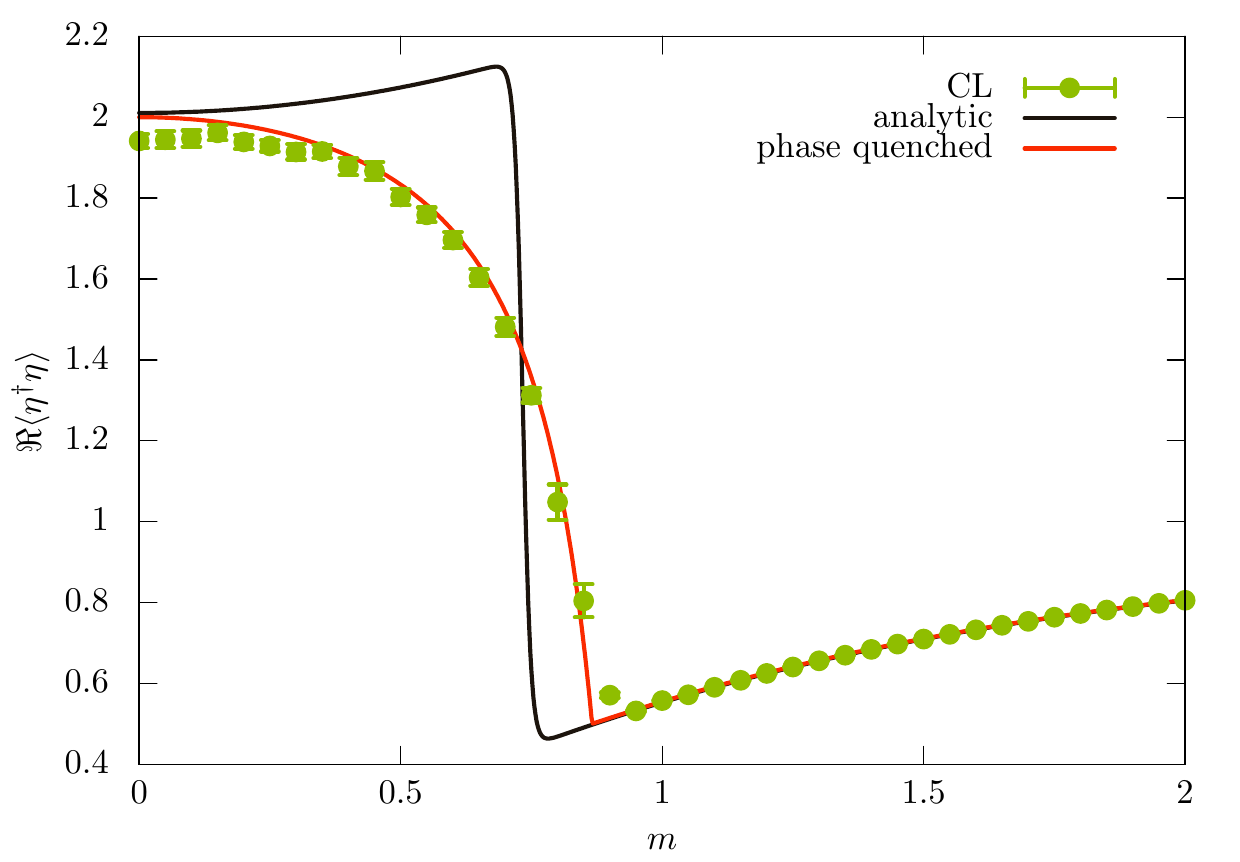} 
  \end{minipage}
  \begin{minipage}[b]{0.5\linewidth}
    \centering
    \includegraphics[width=.95\linewidth]{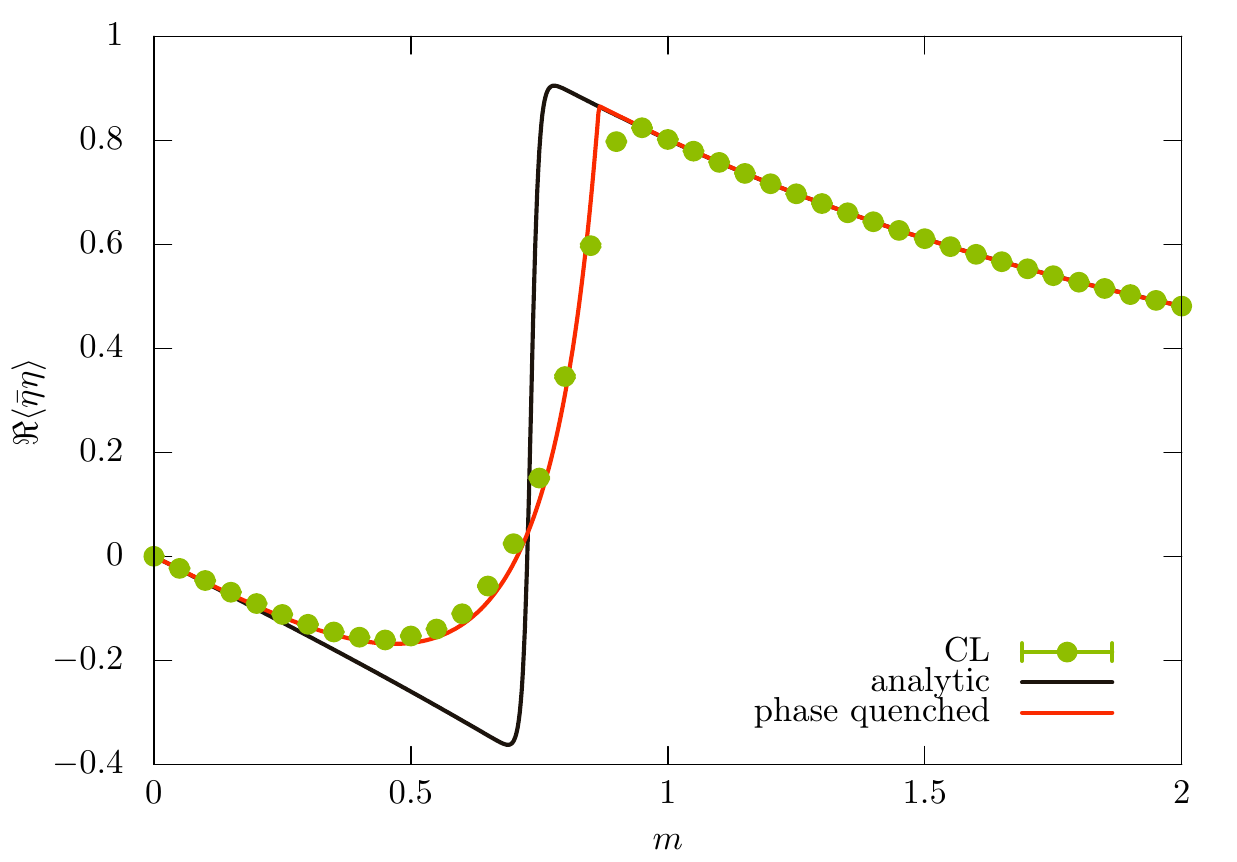} 
  \end{minipage} 
  \caption{\label{mscan} Mass scan for $\mu=0.2$ (upper panel) and $\mu=1$ (lower panel) for matrices with $N=48$. In both cases we plot the baryon number density on the left and the chiral condensate on the right.}
\end{figure}
\FloatBarrier
\section{Reweighted complex Langevin}\label{RCL}
After seeing explicitly that the algorithm fails to reproduce the result of this
complex action model for QCD one naturally asks the question if one could fix
the algorithm in order to enter regions of the parameter space that are initially
inaccessible to CL.
We employ the notion of reweighted complex Langevin (RCL) trajectories~\cite{JB},
that as we will demonstrate, can significantly ameliorate the
convergence properties of the algorithm and lead to trustworthy results
in regions of the parameter space where the naive implementation
of the algorithm was unsuccessful. More precisely we would like to ensure
that the CL does not fail around the transition region.
The main hope is that this reweighting may work better than other conventional forms of reweighting since the
target ensemble with parameters ($\xi=m$, $\mu$) and the auxiliary ensembles
with parameters ($\xi_0=m_0$, $\mu_0)$) have larger overlap. In reweighting one computes the expectation value of an observable $\mathcal{O}$ using
\be
\langle \mathcal{O}\rangle_{\xi}&=&\frac{\int dx \,w(x; \xi) \mathcal{O}(x ;\xi)}{\int dx \,w(x; \xi) }=
\frac{\int dx\, w(x ; \xi_0)\left[ \frac{w(x; \xi)}{w(x ; \xi_0)}\mathcal{O}(x ; \xi)\right]}{\int dx\, w(x ; \xi_0)\left[ \frac{w(x; \xi)}{w(x ; \xi_0)}\right] }
=\frac{\left\langle \frac{w(x; \xi)}{w(x ; \xi_0)}\mathcal{O}(x ; \xi) \right\rangle_{\xi_0}}{\left\langle \frac{w(x; \xi)}{w(x ; \xi_0)} \right\rangle_{\xi_0}} ,
\ee
 but now the weight $w(x ;\xi_0)=e^{-S(x; \xi_0)}$ is complex so one needs to use CL for this ensemble as well. However, this is done via a conservative choice in parameter space where the algorithm converges correctly. As a first case study we choose a relatively small matrix size ($N=6$) and we see in Fig.~\ref{RCLn6} that reweighting can fix all the pathological features of the algorithm. We see that for this small "volume" the analytical answer is completely reproduced for the whole range of parameters. 

\begin{figure}[ht] 
 \begin{minipage}[b]{0.5\linewidth}
    \centering
    \includegraphics[width=.95\linewidth]{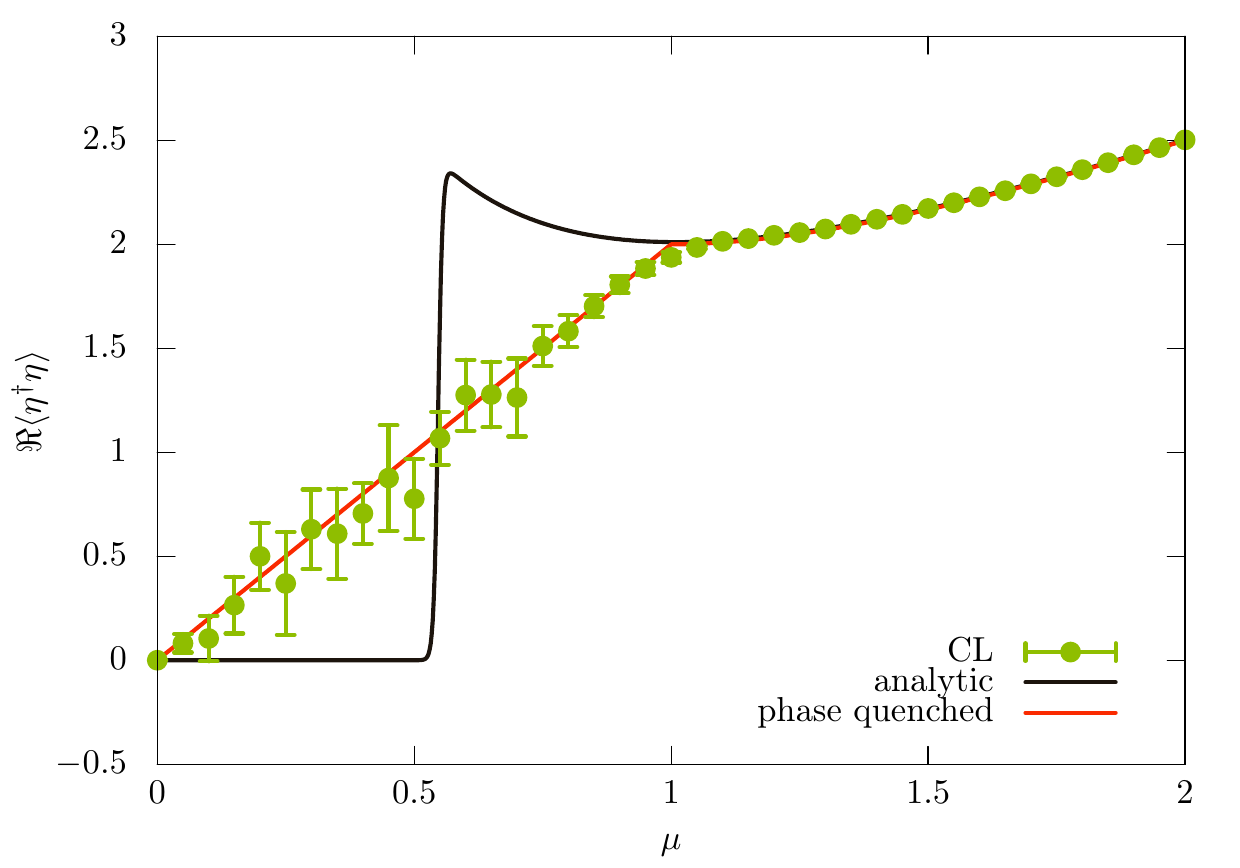} 
    \vspace{4ex}
  \end{minipage}
  \begin{minipage}[b]{0.5\linewidth}
    \centering
    \includegraphics[width=.95\linewidth]{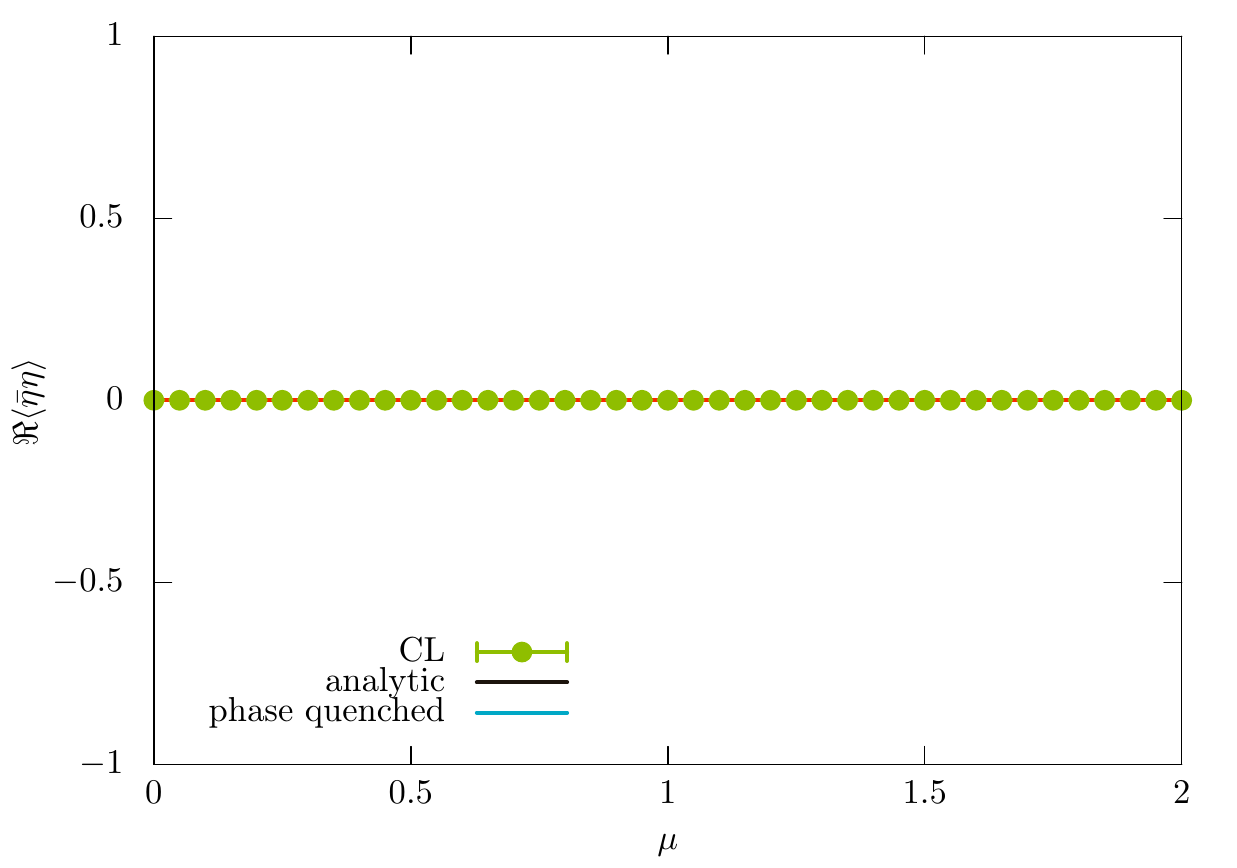} 
    \vspace{4ex}
  \end{minipage} 
  \begin{minipage}[b]{0.5\linewidth}
    \centering
    \includegraphics[width=.95\linewidth]{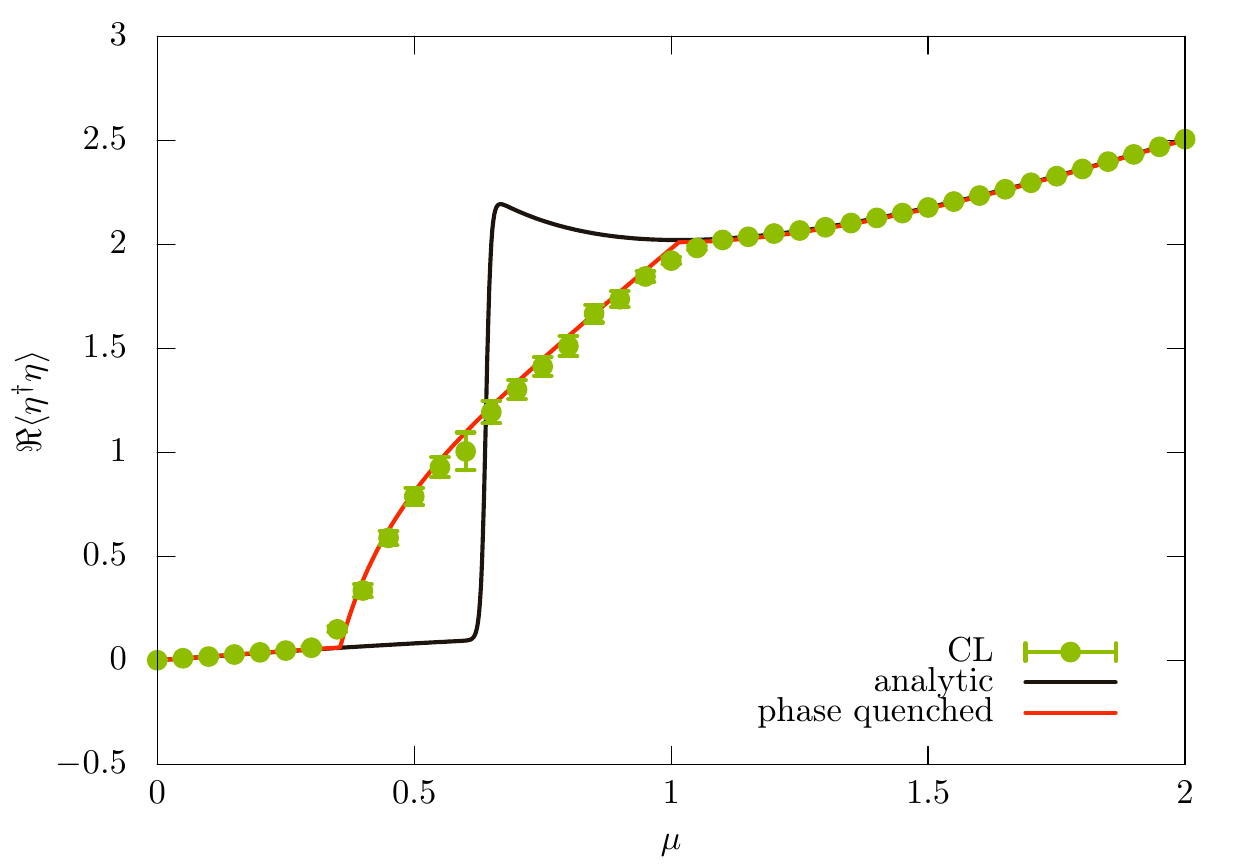} 
    \vspace{4ex}
  \end{minipage}
  \begin{minipage}[b]{0.5\linewidth}
    \centering
    \includegraphics[width=.95\linewidth]{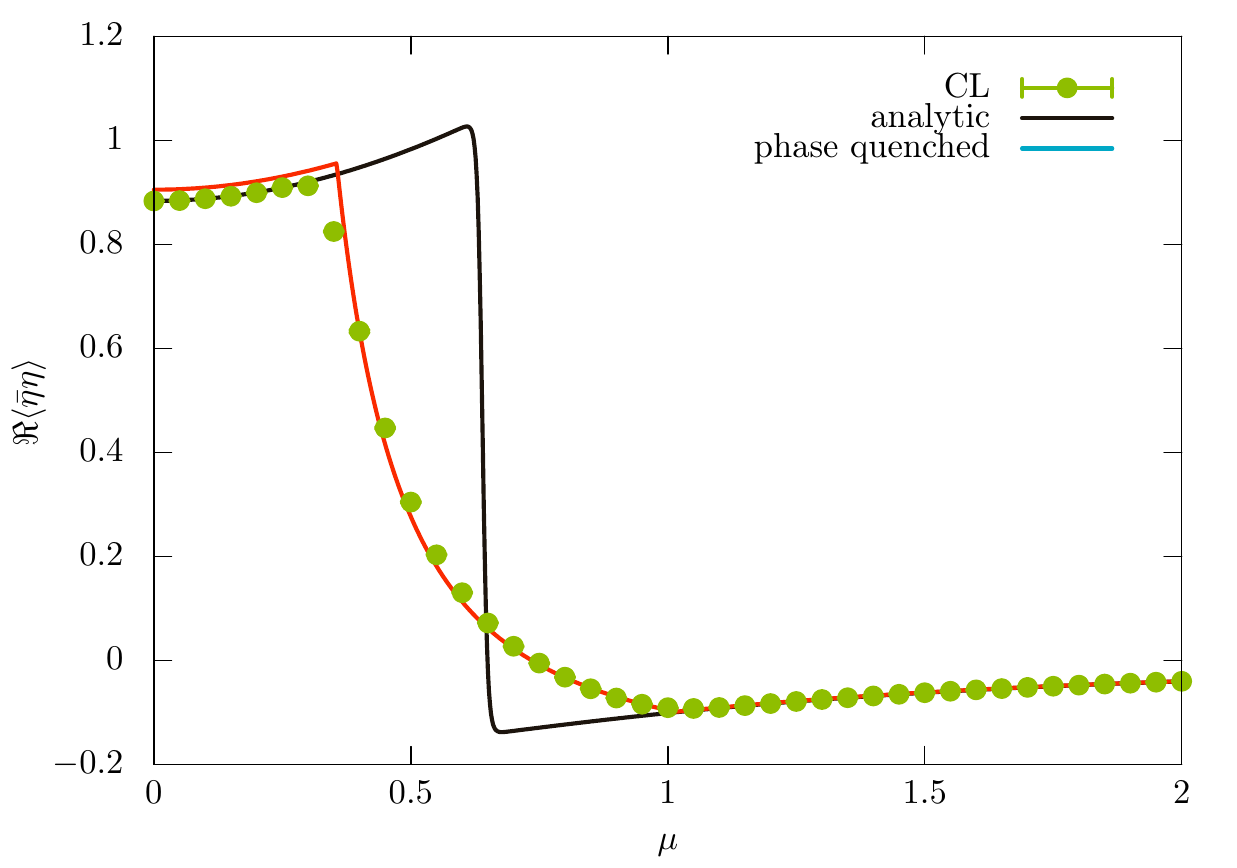} 
    \vspace{4ex}
  \end{minipage} 
  \begin{minipage}[b]{0.5\linewidth}
    \centering
    \includegraphics[width=.95\linewidth]{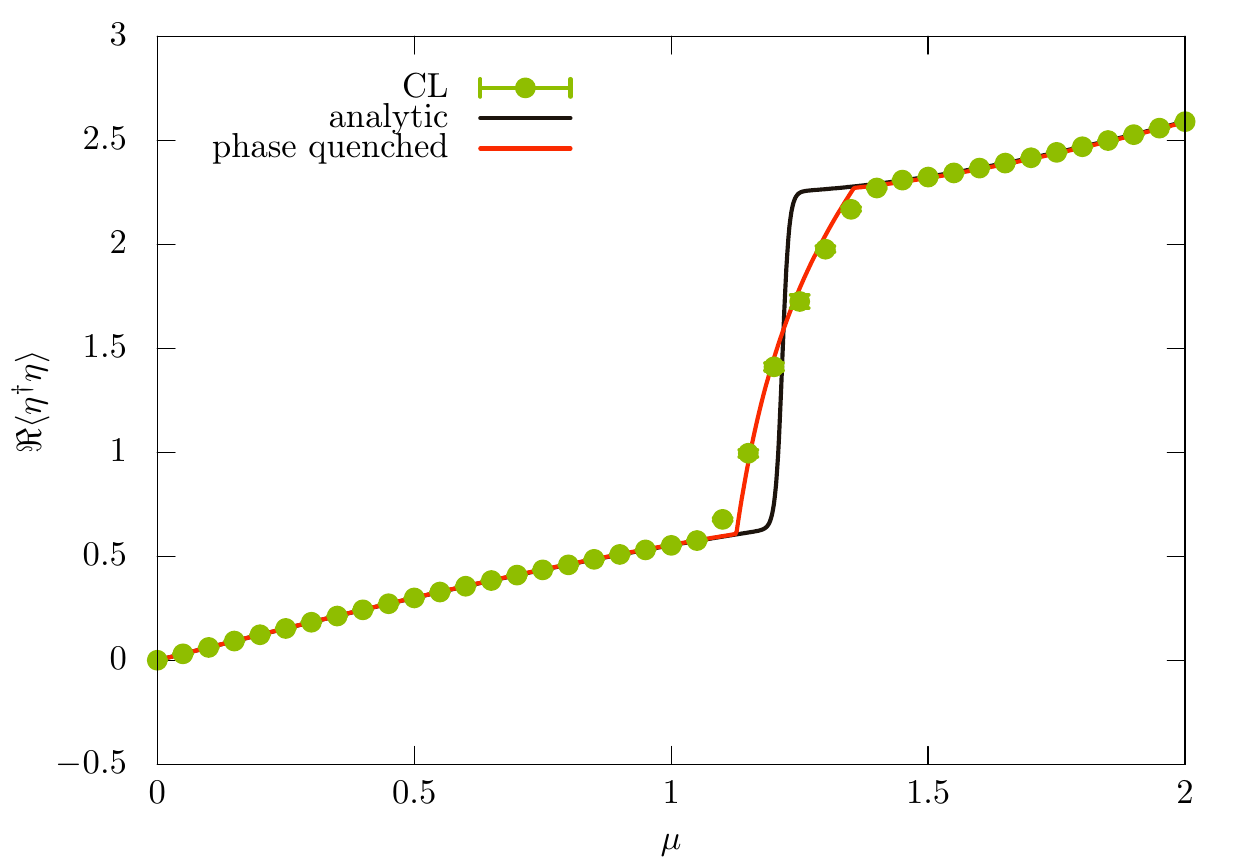} 
  \end{minipage}
  \begin{minipage}[b]{0.5\linewidth}
    \centering
    \includegraphics[width=.95\linewidth]{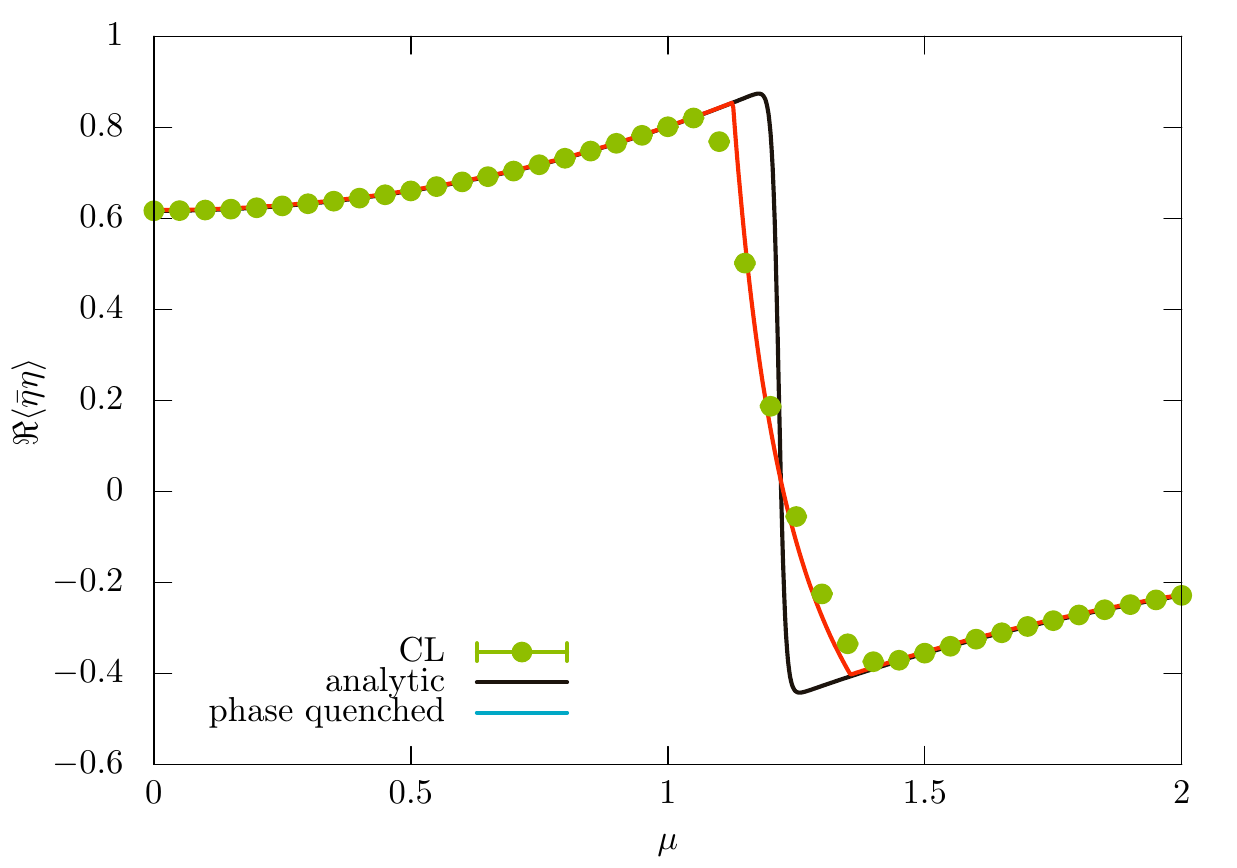} 
  \end{minipage}  
    \caption{  \label{muscan} $\mu$-scan for $m=0$ (upper panel), $m=0.2$ (middle panel) and $m=1$ lower panel for matrices with $N=48$. Again we show the baryon number density on the left and the chiral condensate on the right.}
  \end{figure}
  \FloatBarrier



\begin{figure}[htb]
\centerline
{
\begin{minipage}{0.46\textwidth}
\centerline{%
\hspace{0.1cm}\includegraphics[width=1.07\linewidth]{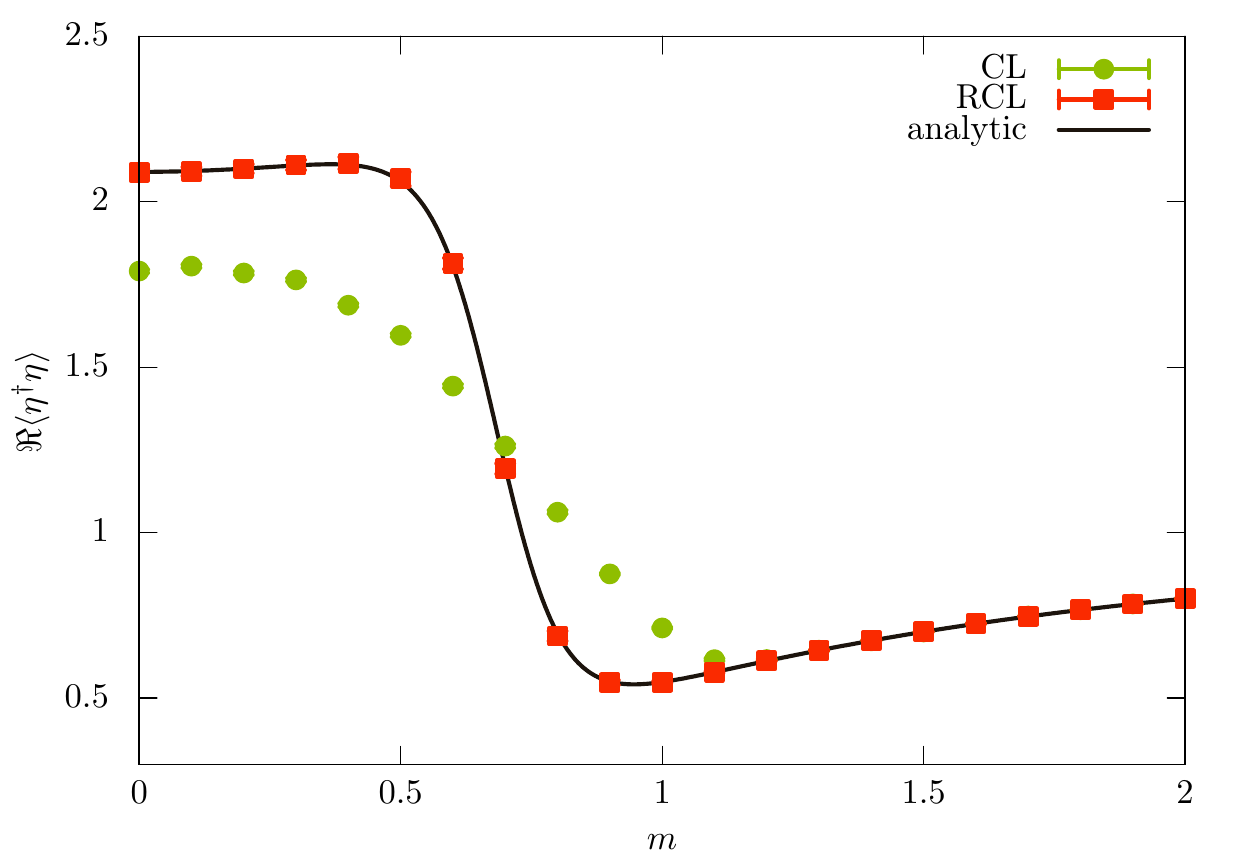}}
\end{minipage}
\hspace{5mm}
\begin{minipage}{0.46\textwidth}
\centerline{%
\hspace{0.5cm}\includegraphics[width=1.07\linewidth]{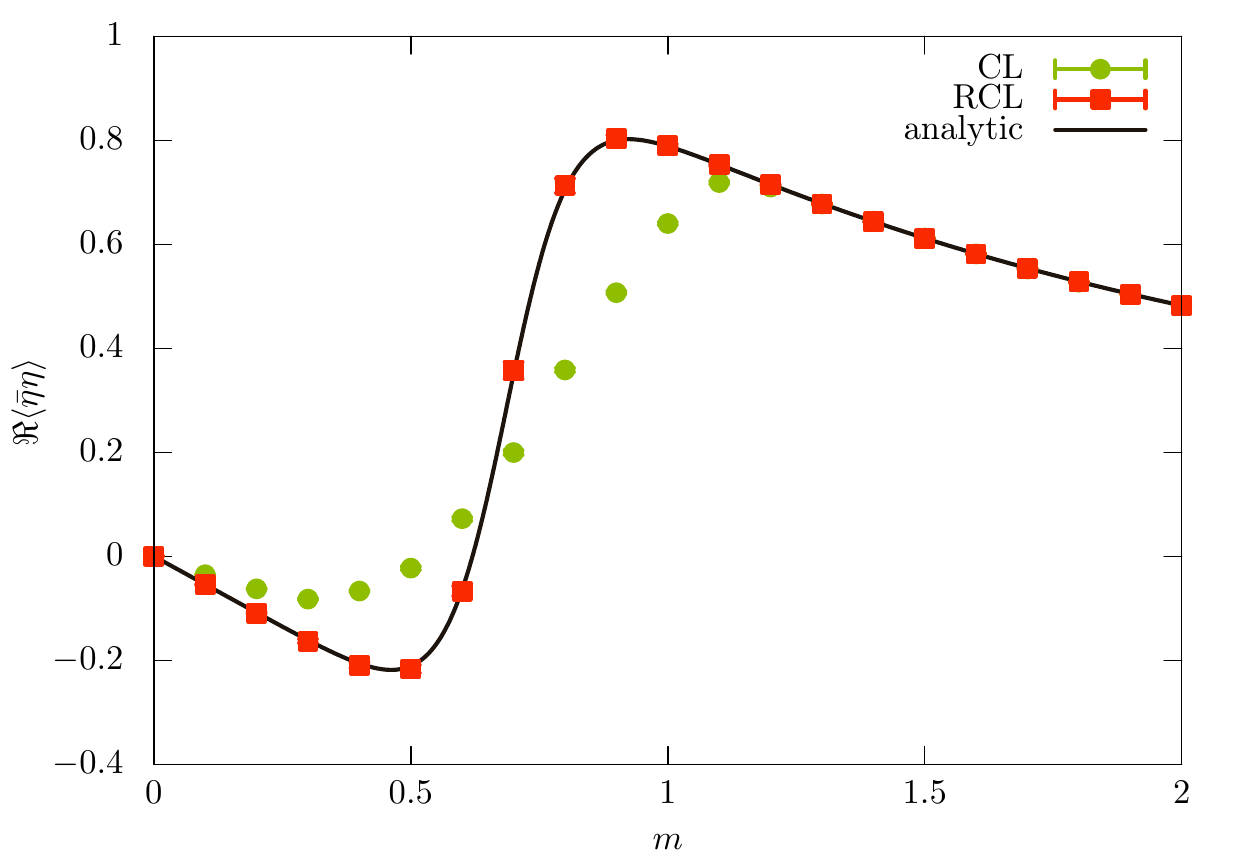}}
\end{minipage}
}
\caption{$\langle \eta^{\dagger}\eta\rangle$ (left) and $\langle \bar{\eta}\eta\rangle$ (right) versus $m$ for $\mu=1$ for matrices with $N=6$ . Results have been reweighted from an auxiliary ensemble with $m_0=4$ and $\mu_0=1$.}
\label{RCLn6}
\end{figure}
\begin{figure}[htb]
\centerline
{
\begin{minipage}{0.46\textwidth}
\centerline{%
\hspace{0.1cm}\includegraphics[width=1.07\linewidth]{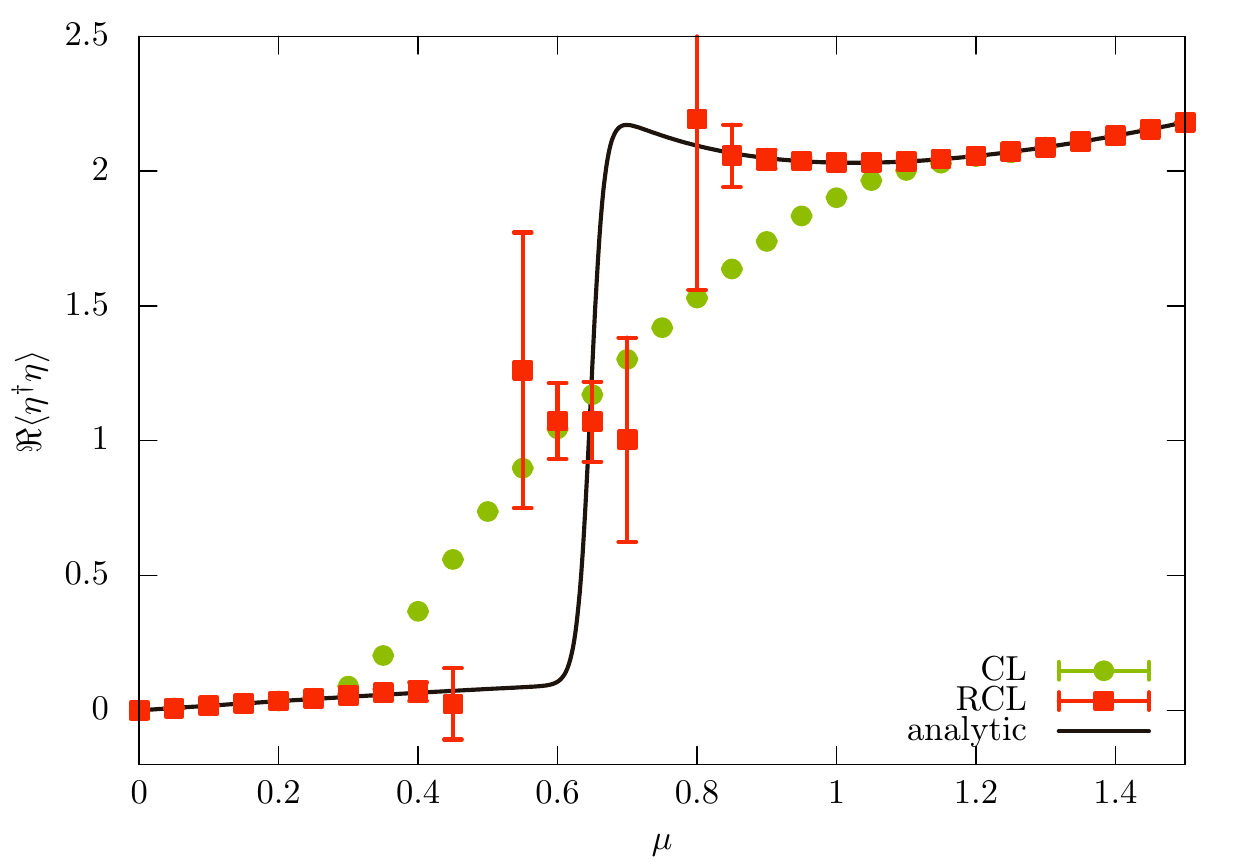}}
\end{minipage}
\hspace{5mm}
\begin{minipage}{0.46\textwidth}
\centerline{%
\hspace{0.5cm}\includegraphics[width=1.07\linewidth]{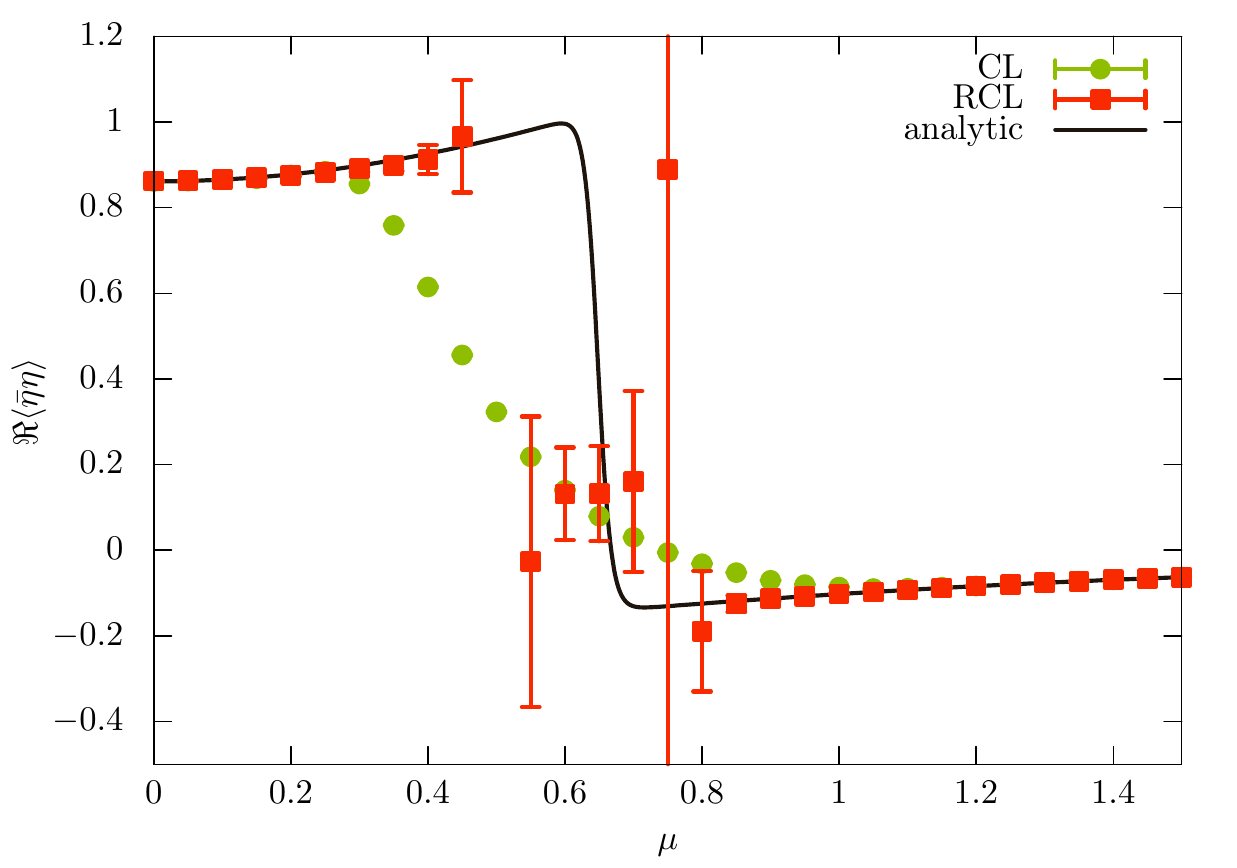}}
\end{minipage}
}
\caption{$\langle \eta^{\dagger}\eta\rangle$ (left) and $\langle \bar{\eta}\eta\rangle$ (right) versus $\mu$  for $m=0.2$ for matrices with $N=24$. Results have been reweighted from an auxiliary ensemble with $m_0=0.2$ and $\mu_0=1.5$.}
\label{RCLn7}
\end{figure}
\begin{figure}[htb]
\centerline
{
\begin{minipage}{0.46\textwidth}
\centerline{%
\hspace{0.1cm}\includegraphics[width=1.07\linewidth]{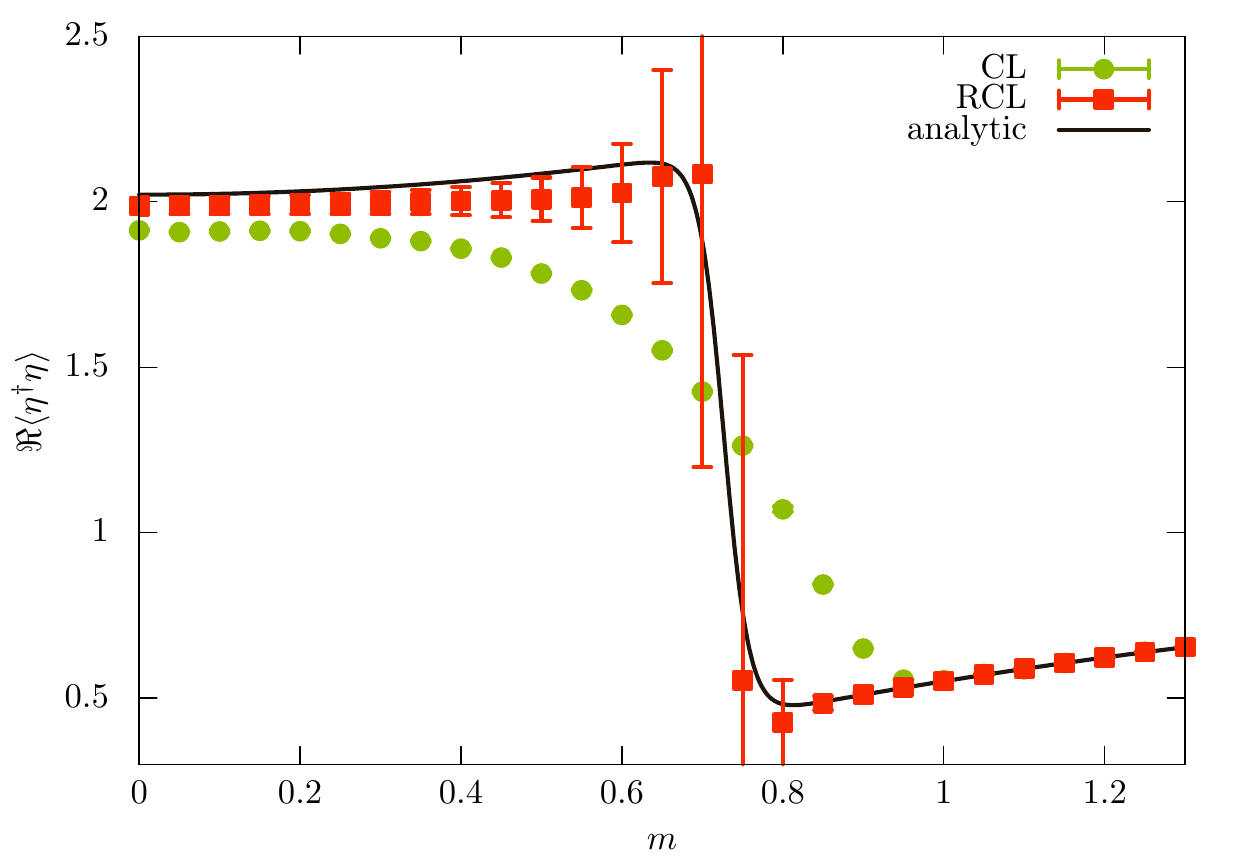}}
\end{minipage}
\hspace{5mm}
\begin{minipage}{0.46\textwidth}
\centerline{%
\hspace{0.5cm}\includegraphics[width=1.07\linewidth]{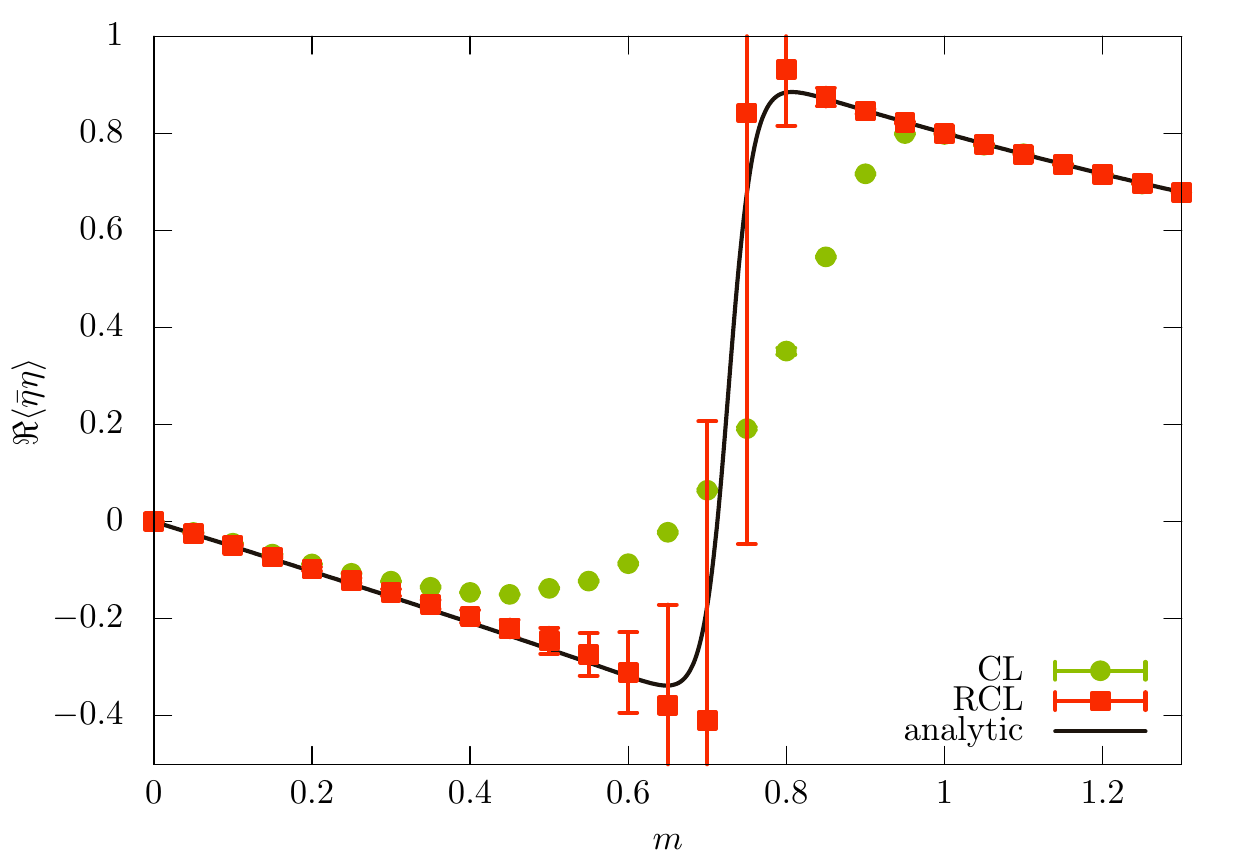}}
\end{minipage}
}
\caption{ $\langle \eta^{\dagger}\eta\rangle$ (left) and $\langle \bar{\eta}\eta\rangle$ (right) versus $m$ for $\mu=1$  for matrices with $N=24$. Results have been reweighted from an auxiliary ensemble with $m_0=1.3$ and $\mu_0=1.0$.}
\label{RCLn8}
\end{figure}

\FloatBarrier
However, we know that the sign problem scales exponentially with the volume and thus one needs to verify that the algorithm is also fixed for larger volumes. We already see in Figs.~\ref{RCLn7} and \ref{RCLn8} that for $N=24$ the agreement with the same statistics as for $N=6$ is worse, but it is interesting to point out that one still has extended the region of correct convergence. 

\begin{figure}[htb]
\centerline
{
\begin{minipage}{0.46\textwidth}
\centerline{%
\hspace{0.1cm}\includegraphics[width=1.07\linewidth]{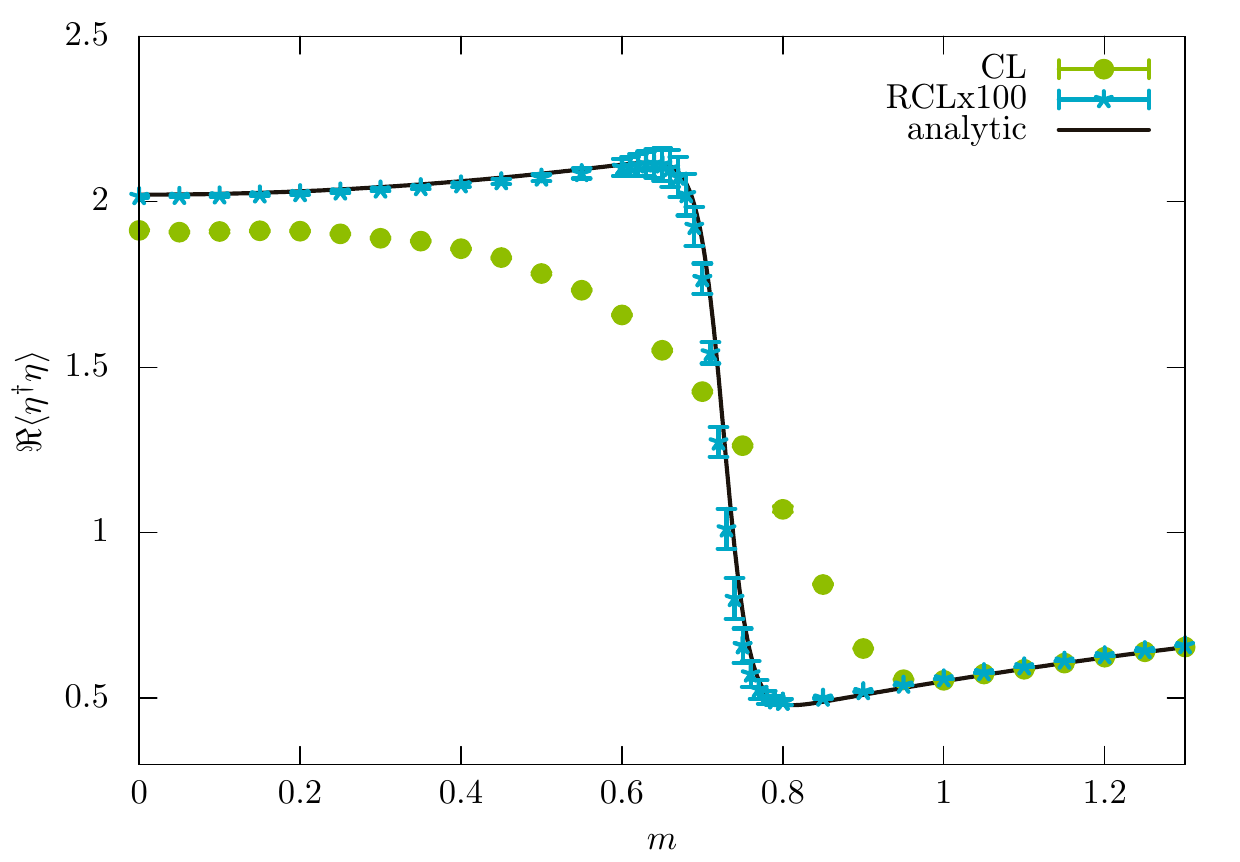}}
\end{minipage}
\hspace{5mm}
\begin{minipage}{0.46\textwidth}
\centerline{%
\hspace{0.5cm}\includegraphics[width=1.07\linewidth]{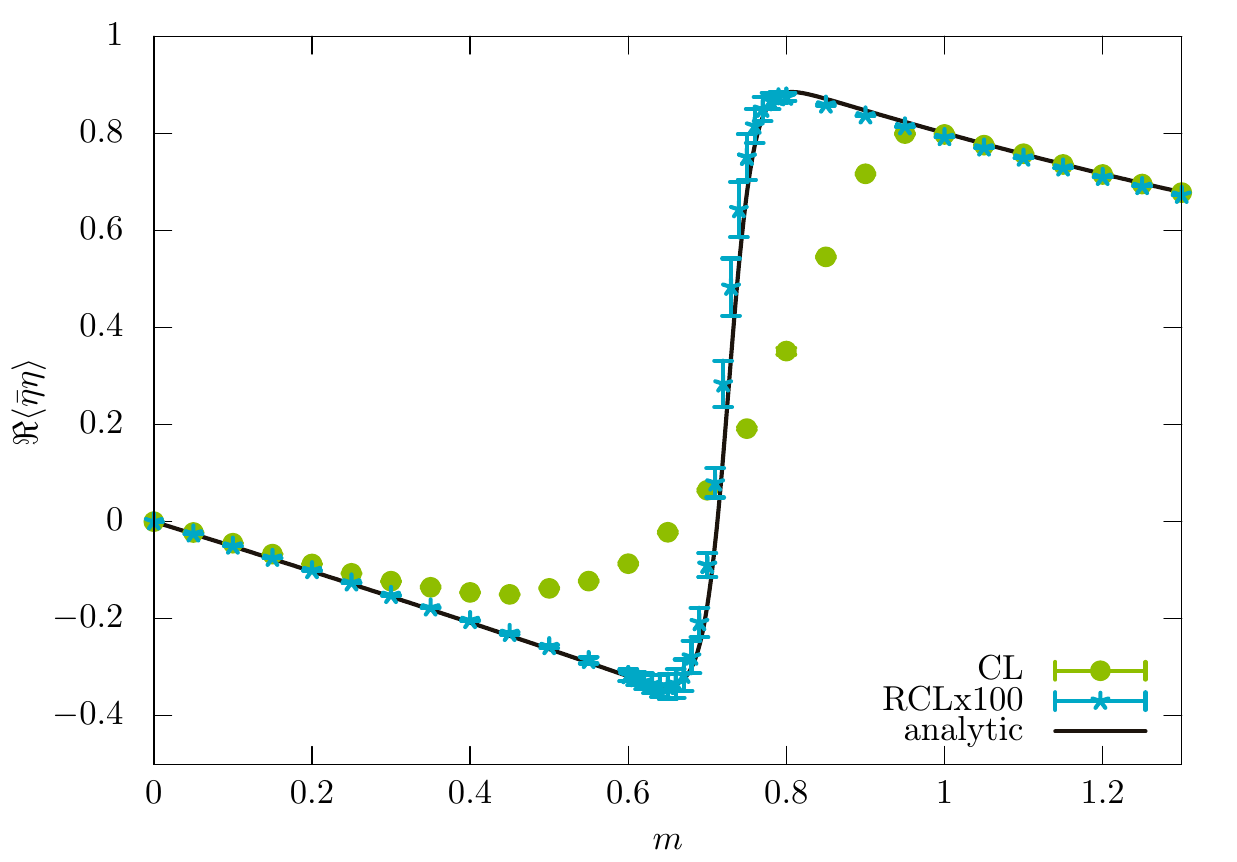}}
\end{minipage}
}
\caption{$\langle \eta^{\dagger}\eta\rangle$ (left) and $\langle \bar{\eta}\eta\rangle$ (right) versus $m$ for $\mu=1$  for matrices with $N=24$. Results have been reweighted from an auxiliary ensemble with $m_0=1.3$ and $\mu_0=1.0$. The RCL results were obtained with 100 times higher statistics than in Fig.~\ref{RCLn8}. }
\label{RCLn9}
\end{figure}

Since this is a case study we would actually try to push the statistics to see if one can obtain the correct analytical result. This is shown in Fig.~\ref{RCLn9} where we have performed a high statistics study by increasing the sample size by a factor of 100 for the mass scan. We see that RCL reproduces the full analytic result in the whole range of parameters.

\FloatBarrier

\section{Conclusions}
In this study we have performed a detailed study of a matrix model for
QCD which possesses a phase transition at a nonzero value of the baryon
chemical potential and we addressed the issue of the convergence
of the complex Langevin algorithm. By direct comparison with the known analytical result
we have observed that the naive implementation of the algorithm converges
to the phase quenched result (whose analytical result in the large-$N$ limit has been first obtained in this study).
We have employed a novel reweighting method in order to cure the pathologies
of the algorithm and we have seen that for small and moderate "volumes" one can
reproduce the full analytical result with high accuracy.
The next logical step, which we wish to address in an upcoming article, is to
find the most efficient reweighting scheme. Once this has been accomplished, and
the results have been verified for a toy model, such as RMT, we plan to put the
algorithm to use on a full QCD simulation.

\section{Acknowledgements} This work was supported by the the Humboldt
Foundation, the National Science Foundation (USA) under grant PHY-1516509 (S.Z.).
JV acknowledges partial support from U.S. DOE Grant  No.\ DE-FAG-88FR40388. JG and OP acknowledge support by the Helmholtz International Center for FAIR within the LOEWE program of the State of Hesse. JB was supported by the Deutsche Forschungsgemeinschaft (SFB/TRR-55). The authors would like to thanks Gert Aarts,  Ion-Olimpiu Stamatescu and Keitaro Nagata for fruitful discussions.

\end{document}